\documentclass{aa}
\usepackage{graphicx}
\usepackage{txfonts}
\usepackage[numbers]{natbib}
\begin{document}

\title{Galactic astroarchaeology: reconstructing the bulge history by means of the newest data}

\author {G. Cescutti\inst{1}
\thanks {email to: cescutti@oats.inaf.it}
\and  F. Matteucci\inst{1,2}}
\institute{Dipartimento di Fisica, Sezione di Astronomia, Universit\'a di Trieste, via G.B. Tiepolo 11, 34131 Trieste, Italy  
\and  I.N.A.F. Osservatorio Astronomico di Trieste, via G.B. Tiepolo 11, 34131 Trieste, Italy}

\date{Received xxxx / Accepted xxxx}

\abstract {The chemical abundances measured in stars of the Galactic
  bulge offer an unique opportunity to test galaxy formation models as
  well as impose strong constraints on the history of star formation
  and stellar nucleosynthesis.}  {The aims of this paper are to
  compare abundance predictions from a detailed chemical evolution
  model for the bulge with the newest data. Some of the predictions
  have already appeared on previous paper (O, Mg, Si, S and Ca) but
  some other predictions are new (Ba, Cr and Ti).}{We compute several
  chemical evolution models by adopting different initial mass
  functions for the Galactic bulge and then compare the results to new
  data including both giants and dwarf stars in the bulge. In this way
  we can impose strong constraints on the star formation history of
  the bulge.}{We find that in order to reproduce at best the
  metallicity distribution function one should assume a flat IMF for
  the bulge not steeper than the Salpeter one. The initial mass
  function derived for the solar vicinity provides instead a very poor
  fit to the data. The [el/Fe] vs. [Fe/H] relations in the bulge are
  well reproduced by a very intense star formation rate and a flat IMF
  as in the case of the stellar metallicity distribution. Our model
  predicts that the bulge formed very quickly with the majority of
  stars formed inside the first 0.5 Gyr.}  {Our results strongly
  suggest that the new data , and in particular the MDF of the 
  bulge, confirm what concluded before and in particular that the
  bulge formed very fast, from gas shed by the halo, and that the
  initial mass function was flatter than in the solar vicinity and in
  the disk, although not so flat as previously thought.   Finally, 
 our model can also reproduce the decrease of the
  [O/Mg] ratio for [Mg/H] $>$ 0 in the bulge, which is confirmed by
  the new data and interpreted as due to mass loss in massive stars.}


\titlerunning{Galactic astroarchaeology of the bulge}

\maketitle

\authorrunning{Cescutti \& Matteucci}

\section{Introduction}

The bulges of spiral galaxies are generally distinguished in true
bulges, hosted by S0-Sb galaxies and ``pseudobulges'' hosted in later
type galaxies (see Renzini 2006 for references). Generally, the
properties (luminosity, colors, line strengths) of true bulges are
very similar to those of elliptical galaxies.  In the following, we
will refer only to true bulges and in particular to the bulge of the
Milky Way.  The bulge of the Milky Way is, in fact, the best studied
bulge and several scenarios for its formation have been put forward in
past years.  As summarized by Wyse \& Gilmore (1992) the proposed
scenarios are: i) the bulge formed by accretion of extant stellar
systems which eventually settle in the center of the Galaxy.  ii) The
bulge was formed by accumulation of gas at the center of the Galaxy
and subsequent evolution with either fast or slow star formation.
iii) The bulge was formed by accumulation of metal enriched gas from
the halo or thick disk or thin disk in the Galaxy center.  
More recently, Elmegreen (2008, 2009) has proposed a bulge formation due to
coalescence of large clumps in primordial galaxies. These clumps form
by gravitational instability. The model requires fast gas assembly and
not a hierarchical merging of pre-existing star-rich galaxies as in
some hierarchical build-up models. Therefore, this model suggests a
fast assembly of bulges.

The metallicity distribution function (MDF) of stars in the bulge and
the [$\alpha$/Fe] ratios greatly help in selecting the most probable
scenario for the bulge formation.  Matteucci \& Brocato (1990)
suggested that in order to fit the MDF of the bulge one should assume
that it formed very quickly in less than 1 Gyr and as a consequence of
this they predicted that the [$\alpha$/Fe] ratios in the bulge stars
should be oversolar for a large interval of [Fe/H].  Their suggestion
was based on the assumption that galaxies of different morphological
type suffer different histories of star formation: in particular,
ellipticals and true bulges should experience a strong burst of star
formation lasting for a short time, whereas spirals and even more
irregulars suffer a milder and continuous star formation rate
(SFR). These assumptions, coupled with the time-delay between the Fe
enrichment from Type Ia supernovae (white dwarfs in binary systems)
and the $\alpha$-element enrichment from core collapse supernovae
(originating in massive stars, Type II, Ib/c), produce a different
behaviour of the [$\alpha$/Fe] vs. [Fe/H] relations under different
SFRs. In other words, objects such as bulges evolve very fast and the
[$\alpha$/Fe] ratios in the majority of their stars are expected to be
high and oversolar for a large interval of [Fe/H].  This is due to the
fact that, since star formation is very intense, the bulge reaches
very soon a solar metallicity thanks only to the core collapse SNe,
which produce some Fe, then when SNe Ia, which produce the bulk of Fe,
start exploding the change in the slope occurs at a larger [Fe/H]
($\sim$ 0) than in the solar vicinity ($\sim$ -1.0). For spiral and
irregular instead the Fe enrichment by core collapse SNe is much less,
due to the milder SFR, and when Type Ia SNe start appearing the [Fe/H]
is still low. Thus the change in slope in irregulars occurs at [Fe/H]
$\le$ -2.0 (see Matteucci, 2001).

Years later, McWilliam \& Rich (1994) observed some [$\alpha$/Fe]
ratios in stars in the Baade's window and concluded that [Mg/Fe] was
indeed high for a large range of [Fe/H]. For other elements, such as
oxygen, the situation was not so clear. These data were derived from
low resolution spectra. Minniti (1996) concluded from kinematics and metallicities
of red giant stars in the field that the bulge formed quickly by dissipative collapse, from material left over 
after the formation of the halo.  
In the following years a great deal of 
observations of bulge stars appeared: high resolution abundances were
derived by Zoccali et al. (2006), Fulbright et al. (2007),  Lecureur et
al. (2007) and these studies suggested an oversolar and almost
constant value for a large [Fe/H] range also for other
$\alpha$-elements, such as O, thus confirming that the abundance
ratios indicate a fast bulge formation.  From the theoretical point of
view, Ballero et al. (2007a, hereafter BMOR07)) presented an updated
model relative to Matteucci \& Brocato (1990) for the bulge. This
model includes stellar feedback and the development of a galactic wind
which occurs when most of star formation in the bulge is over.  Very
detailed predictions were given in this paper for several elements and
for the MDF. The agreement with observations was good suggesting that
the bulge formed on a time scale between 0.3 and 0.5 Gyr, that the star
formation was much more efficient than in the solar vicinity (by a
factor of $\sim$ 20) and that an IMF much flatter than the Scalo
(1986) or Kroupa et al. (2001), adopted for the solar neighbourhood,
was required. In particular, this is required by the observed MDF.
Recently, many more abundance data for bulge and thick disk giant
stars appeared Alves-Brito et al. 2010) measuring several
$\alpha$-elements and Fe.  The observed trends confirm the previous
papers and found a similarity between the thick disk and bulge
stars. Moreover, Bensby et al. (2010) and Johnson et al. (2007, 2008) for
the first time measured the same abundance ratios in microlensed dwarf
and subgiant stars in the bulge and found good agreement with the
abundances measured in giant stars.

The paper is organized as follows: in Section 2 we describe the
observational data, in Section 3 we briefly describe the chemical
model for the bulge, in Section 4 we compare our model results with
the newest data and finally in Section 5 we summarize our conclusions.

\section{Observational data}
In this work, we choose to use only the most recent observational data for the bulge.
We select the chemical abundance data calculated by Bensby et al. (2010)
Alves-Brito et al. (2010) e Ryde et al. (2009).
The spectra are obtained by the different authors using different techniques:
Bensby et al. (2010) perform a detailed elemental abundance analysis of dwarf stars in the Galactic bulge, 
based on high-resolution spectra
that were obtained while the stars were optically magnified during gravitational microlensing events;
Alves-Brito et al. (2010) use high-resolution optical spectra of 25 bulge giants in Baade's window 
and 55 comparison giants (4 halo, 29 thin disk and 22 thick disk giants) in the solar neighborhood;
Ryde et al. (2009) obtain high-resolution, near-infrared spectra 
in the H band are recorded using the CRIRES spectrometer on the Very 
Large Telescope, the
CNO abundances can all be determined from the numerous molecular lines in the 
wavelength range observed, abundances of the $\alpha$ elements
Si, S, and Ti are also determined from the near-IR spectra.
For a comparison with our previous work, we decide to show for 
the  elements Mg and O the data selected in Cescutti et al. (2009).
These data are fully described in McWilliam et al. (2008).

We compare our model results for the MDF 
in the bulge 
with the MDF determined by Zoccali et al. (2008). 
They observed  about 800 bulge field K giants with the GIRAFFE spectrograph 
of FLAMES at the VLT  at spectral resolution $R\sim$ 20 000. 
The iron abundances, resulting of their LTE analysis,
allowed to construct a MDF for the bulge that, for the first time, 
is based on high-resolution spectroscopy for each individual star.

\section{The chemical evolution model}

The adopted basic chemical evolution model closely follows that
in BMOR07. 
The main assumption is that the Galactic bulge formed by the
fast collapse of primordial gas (the same gas out of which the
halo was formed) accumulating in the center of our Galaxy. 
We recall the fundamental ingredients of this model:
\begin{itemize}
\item[-] Instantaneous mixing approximation: the gas over the whole
  bulge is homogeneous and well mixed at any time.
\item[-] Star formation rate (SFR) parameterized as follows:
\begin{equation}
\psi(r,t) =
\nu G^k(r,t)
\end{equation}
where $\nu$ is the star formation efficiency (i.e. the inverse of the
timescale of star formation) in the bulge, $k = 1$ is chosen to
recover the star formation law employed in models of spheroids
(e.g. by Matteucci, 1992) and $G(r,t) =
\sigma_{gas}(r,t)/\sigma(r,t_G)$ is the normalized gas surface mass 
density (where $\sigma_{gas}(r,t)$ is the gas surface mass density
and $\sigma(r,t_G)$ is the surface gas density of the bulge at the
present time $t_G=13.7$ Gyr). 
\item[-] The initial mass function (IMF) is expressed as a power
  law with index $x$:
\begin{equation}
\phi(m) \propto m^{-(1+x)}
\end{equation}
within the mass range $0.1-100M_{\odot}$.  In this paper we adopt as a
reference model the IMF suggested by BMOR07, namely a two slope IMF
with $x=0.95$ for $M > 1 M_{\odot}$ and $x=0.33$ for $M < 1
M_{\odot}$, and we test also a one slope Salpeter (1955) IMF (x=1.35)
and the two-slope Scalo (1986) IMF.

\item[-] The gas which  forms the bulge has a primordial chemical
  composition and the accretion rate is given by:
\begin{equation}
\dot{G}(r,t)_{inf}=\frac{A(r)}{\sigma(r,t_G)}e^{-t/\tau}
\end{equation}
where $\tau$ is an appropriate collapse timescale and $A(r)$ is
constrained by the requirement of reproducing the current total
surface mass density in the Galactic bulge. 
Actually, we should use the halo chemical composition for the
infalling gas, but our simulations have demonstrated that unless very high
$\alpha$-enhancements are adopted, the results are essentially the
same.
\noindent\item[-] The instantaneous recycling approximation is relaxed;
  stellar lifetimes are taken into account in detail following the
  prescriptions of Kodama (1997).
\noindent\item[-] Detailed nucleosynthesis prescriptions are taken from: 
      i) Fran\c cois et al. (2004), who made use of widely adopted
      stellar yields and compared the results obtained by including
      these yields in a detailed chemical evolution model with the
      observational data, with the aim of constraining the stellar
      nucleosynthesis.  For low- and intermediate-mass ($0.8 -
      8M_{\odot}$) stars, which produce $^{12}$C, N and heavy
      $s$-elements, yields are taken from the standard model of Van
      den Hoek \& Groenewegen (1997) as a function of the initial
      stellar metallicity. Concerning massive stars ($M>10M_{\odot}$),
      in order to best fit the data in the solar neighbourhood, when
      adopting Woosley \& Weaver (1995) yields, Fran\c cois et
      al. (2004) found that O yields should be adopted as a function
      of the initial metallicity, Mg yields should be increased in
      stars with masses $11-20M_{\odot}$ and decreased in stars larger
      than $20M_{\odot}$, and that Si yields should be slightly
      increased in stars above $40M_{\odot}$; as in BMOR07 we use here
      their constraints on the stellar nucleosynthesis to test whether
      the same prescriptions give good results for the Galactic bulge,
      when compared with the newest data.  ii) In the range of massive
      stars we have also adopted the yields of Maeder (1992) and
      Maeder \& Meynet (2002) containing mass loss.  The effect of
      mass loss is visible only for metallicities $\ge Z_{\odot}$.
      The use of these yields is particularly important for studying
      the evolution of O and C, the two most affected elements (see
      McWilliam et al. 2008; Cescutti et al. 2009). iii) For  Ba,
      we use the nucleosynthesis prescriptions adopted by Cescutti et al. 
      (2006) to best 
      fit the observational data for this neutron capture element 
      in the solar vicinity; the same nucleosynthesis prescriptions give also
      good results when applied to dwarf spheroidals (Lanfranchi et al. 2006)
      and to the Galactic halo using an inhomogeneous model  (Cescutti 2008).
      In particular, we assume that the the s-process fraction of 
      Ba is produced in low mass stars ($1-3M_{\odot}$), whereas the r-process fraction 
     of Ba originates from stars in the range $12-30M_{\odot}$.

\noindent\item[-] The Type Ia SN rate was computed according to Greggio \&
  Renzini (1983) and Matteucci \& Recchi (2001). 
  Yields are taken from Iwamoto et al. (1999) which is an updated
  version of model W7 (single degenerate) from Nomoto et al. (1984).
  These supernovae are the main contributors of Fe and produce small
  amounts of light elements; they also contribute to some extent to
  the enrichment in Si and Ca.

\noindent\item  The treatment of the supernova-driven galactic wind is the same as in BMOR07, 
where we address the reader for details.
 \end{itemize}

\begin{figure}
\begin{center}
\includegraphics[width=.49\textwidth]{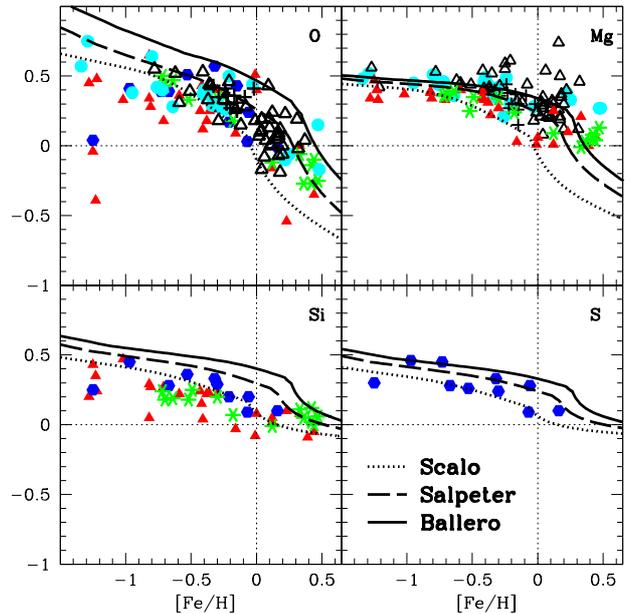}

\caption{ Comparison between the predictions of our  model using 3 different 
IMF for [O/Fe],[Mg/Fe],[Si/Fe] and [S/Fe] vs [Fe/H] and the observations in the bulge.
The observational data for the bulge are:
the filled circles from MFR09 and FMR07;
the  open triangles from  Lecureur et al. (2007);
the filled triangles from Alves-Brito et al. (2010);
the plus signs are the infrared results from Rich \& Origlia (2005);
the filled hexagons are the infrared results by Ryde et al. (2009);
and the stars are the results for microlensed dwarf stars 
by Bensby et al. (2010).}\label{fig1}

\end{center}
\end{figure}

\section{Results}
\subsection{Constraints on the IMF}
We assumed, as in previous papers, a high efficiency of star formation
($\nu=20 Gyr^{-1}$) and a short timescale for gas accretion ($\tau=0.5
Gyr$). It is supposed that the gas which formed the bulge was that
shed by the halo and therefore it was slightly enriched in
metallicity. The values of the above parameters produce a very fast
bulge formation, as required to reproduce the bulge abundance data.
We have computed several models for the Galactic bulge by varying the
IMF: in particular we adopted i) the very flat IMF as described in
BMOR07, ii) the Salpeter (1955) IMF and iii) the Scalo (1986) IMF.  In
Figure 1 we show the predicted [$\alpha$/Fe] ratios versus [Fe/H] as
predicted and observed in the bulge.

\begin{figure}
\begin{center}
\includegraphics[width=.49\textwidth]{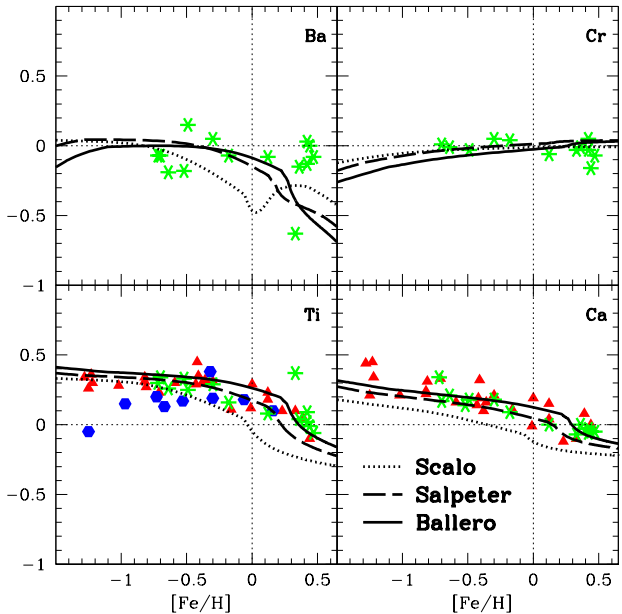}
\caption{ Comparison between the predictions of our  model using 3 different 
IMF for [Ba/Fe], [Cr/Fe], [Ti/Fe] and [Ca/Fe] vs [Fe/H]. The symbols for the
observational data are the same as in Fig. 1.}\label{fig2}
\end{center}
\end{figure}

This kind of diagram is usually interpreted as due to the time-delay
between the chemical enrichment from the Type II and the Type Ia
SNe. Another fundamental parameter in this graph is the SFR which
determines the age-abundance relations and the shape of the
[$\alpha$/Fe] vs. [Fe/H] relations (see Matteucci 2001). In
particular, the long plateau and the oversolar values observed for the
[$\alpha$/Fe] ratios in bulge stars extending to solar metallicity, is
well reproduced by a very fast bulge formation.  As one can see,
different IMFs predict different absolute [$\alpha$/Fe] ratios, in the
sense that flatter is the IMF and higher are the [$\alpha$/Fe] ratios,
since in a flatter IMF there are relatively more massive stars than in
a steeper one.  In addition, different IMFs tend to produce small
variations on the knee of the [$\alpha$/Fe] ratios, in the sense that
a flatter IMF predicts a knee of [$\alpha$/Fe] ratio at larger [Fe/H]
values. Therefore, the length of the plateau can in principle be 
used to impose constraints on the IMF.    However, 
for the bulge the knees occur always at [Fe/H]$>$ 0
since the strong assumed SFR always induces a very fast increase of
the [Fe/H], thus having the SNe Ia occurring when the ISM has already
reached a solar Fe abundance. It is worth noting that the effect of
varying the IMF is different for different elements, being more
evident for O and almost negligible for S. This of course depends on
the specific progenitors of each elements and in particular on whether
two elements are produced by the same stars or in different mass
ranges: the largest difference is, in fact, seen in the O plot, since
O is mainly produced in massive stars whereas Fe is mainly produced in
low and intermediate mass stars (SNe Ia). In the case of S and Si
instead, SNe Ia contribute in a non-negligible way to these two
elements. From a look at Figure 1 we can conclude that the Scalo IMF
predicts too low [$\alpha$/Fe] ratios, whereas the Salpeter and BMOR07
IMFs produce results more in agreement with observations.  In Figure 2
we show the predicted and observed [el/Fe] vs. [Fe/H] for el=Ba, Cr,
Ti and Ca.  This is the first time that predictions for Ba, Cr and Ti
for the bulge are presented. Also in this case the general agreement
with the data is good and the best one is for the BMOR07 and Salpeter
IMF, thus reinforcing the conclusion that the IMF in the bulge should
be flatter than in the disk, as claimed by many papers before
(e.g. Matteucci \& Brocato, 1990; Matteucci et al. 1999; BMOR07).
This result has important implications for the [$\alpha$/Fe] ratios
and their behaviour with [Fe/H] in bulge and thick disk stars
(Chiappini et al. in preparation). In fact, there are some indications
(e.g. Alves-Brito \& al. 2010; Mel\'endez \& al. 2008) that some ratios
are the same for bulge and thick disk stars.

\begin{figure}
\begin{center}
\includegraphics[width=.49\textwidth]{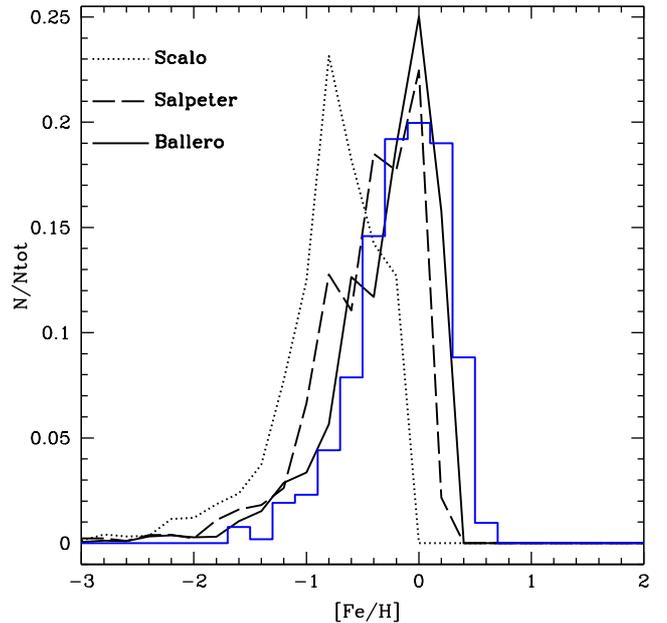}
\caption{ Comparison between the predictions for the MDF of our  model 
using 3 different IMF and MDF measured by Zoccali et al. (2008) 
in the bulge stars; the observed MDF is represented by an histogram.}\label{MDF}
\end{center}
\end{figure}

However, the most convincing evidence for an IMF flatter in the bulge
is provided by Figure 3 where the predicted and observed stellar
metallicity distribution function (MDF) for bulge stars is plotted. As
one can see, the position of the peak in the bulge MDF is extremely
sensitive to the assumed IMF.  A Scalo IMF, which is good for
reproducing the solar neighbourhood properties, it fails completely
for the bulge.  
The results obtained by Zoccali et al. (2008) are also consistent 
with the presence of a gradient in the bulge, as the ones by Minniti (1996).
So, their findings support our scenario in which both infall and outflow are important 
and in which the bulge formed very fast by dissipative collapse of gas
shed by the halo, rather than the scenario in which the bulge would result solely
from the vertical heating of the bar. On the other hand, the presence
of a radial metallicity gradient warns us about the limitations of our model which predicts
only a global MDF for the bulge.

\subsection{Constraints on the stellar nucleosynthesis}
\begin{figure}
\begin{center}
\includegraphics[width=.49\textwidth]{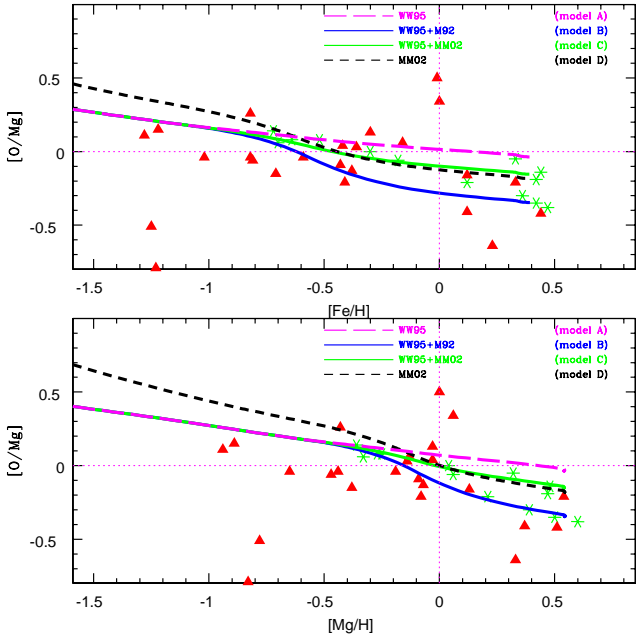}
\caption{ Comparison between the predictions of our 4 models for [O/Mg] vs [Mg/H] (upper panel)
and [O/Mg] vs [Fe/H] (lower panel). The symbols for the
observational data are the same as in Fig. 1}\label{omgmgh}
\end{center}
\end{figure}

A very important result is the change in slope observed for bulge
stars in the ratio [O/Mg] for [Mg/H]$>0$. McWilliam et al. (2008) and
Cescutti et al. (2009) interpreted this effect as due to the mass loss
in massive stars. In fact, the mass loss becomes important only for
metallicities larger than solar and induces the effect of producing
more C and He, which are the elements preferentially lost through mass
loss by stellar winds, at expenses of O which is produced in a lower
amount due to the loss of its progenitor elements, C and He.  This
effect therefore involves only C and O among heavy elements, thus
producing a lowering of the [O/Mg] ratio for [Mg/H] $>$ 0. In this
paper we show again the [O/Mg] vs. [Mg/H] and vs. [Fe/H] (Figure 4)
compared also with data relative to dwarf stars in the bulge. As one
can see, also the dwarf stars confirm the change in slope for the
[O/Mg] ratio. In this figure are shown the predictions for models with
different nucleosynthesis prescriptions (see Cescutti et al. 2009 for
details): a model with massive stars evolving without mass loss (model
A) which shows no change in the slope, as expected; a model with the
mass loss prescriptions from Maeder (1992) for $Z> Z_{\odot}$ (model
B) which gives the best agreement with the observations; two models
including the more recent prescriptions for mass loss by Meynet \&
Maeder (2002): model C  only for  $Z> Z_{\odot}$  and model D 
for all the metallicities. It is worth noting that the same
figures are shown by Alves-Brito et al. (2010) who had the files of
our models: however, they plotted twice [O/Mg] vs. [Fe/H] instead of
[O/Mg] vs. [Fe/H] and [O/Mg] vs. [Mg/H] and then concluded that the
plot as a function of [Mg/H] did not provide a good fit. Here we show
the correct file and the agreement with the data is indeed very good.

Therefore, we confirm here that also the newest bulge data, including
dwarf stars, indicate a change in slope in the [O/Mg] and that this is
best explained by mass loss in massive stars which becomes important
for metallicities larger than solar.

\section{Discussion and Conclusions}

In this paper we have shown a comparison between the predictions of a
model for the Galactic bulge concerning chemical abundances and the
most recent abundance data. The adopted model is that of BMOR07
assuming a very fast SFR with an efficiency 20 times higher than
assumed for the solar vicinity and an infall timescale of 0.3 Gyr. It
is in fact assumed that the bulge formed by accretion of material lost
from the the halo.  In particular, some of the predictions (O, Mg, Si,
S and Ca) already appeared on previous papers but here they are
compared to different and newer data, whereas other predictions
concerning Ba, Cr and Ti are new. We tested different IMFs (Salpeter,
1955; Scalo 1986; BMOR07).  Our main conclusions can be summarized as
follows:

\begin{itemize}
\item In order to reproduce the MDF of the bulge stars an IMF flatter 
 than the Scalo (1986) one should be assumed.
 Note that the Scalo IMF, as well as the Kroupa et al. (1993) and 
Kroupa (2001) IMF which are  good for the solar vicinity do not fit the MDF. 
This result is very  important because it suggests that the IMF in the bulge was different
than in the disk. The best IMF is the very flat one suggested by
BMOR07. The same conclusion was reached by Ballero et al. (2007b) who
analyzed the IMF of Kroupa (2001).

\item In order to best fit the [el/Fe] vs. [Fe/H] relations for O, Mg,
  Si, S, Ti, Cr and Ba an IMF flatter than the Scalo one is also
  required. In these cases a Salpeter IMF can also be
  acceptable. Again, the Scalo and similar IMFs should be
  rejected. These abundance patterns suggest that the bulge formed
  very quickly and that the majority of bulge stars were in place
  already in the first 0.5 Gyr.

\item We compared for the first time theoretical predictions with data 
from dwarf stars in the bulge. These new data agree with the previous
ones relative to bulge giants and agree with our predictions. In
particular, the dwarf data confirm the the change in slope in the
[O/Mg] vs. [Mg/H] observed for [Mg/H] $>$0, a trend which has been
already successfully interpret as due to mass loss in massive stars. In
particular, a model including stellar yields computed with a high rate
of mass loss, such as that suggested by Maeder (1992), still gives the
best agreement with data.

\item We need a flatter IMF than in the the solar vicinity but 
 the novelty relative to the BMOR07 paper is that we found that  
 a Salpeter IMF with x=1.35 over the whole stellar mass range can also
reproduce all the bulge data without the need of invoking an extremely
flat IMF, as suggested in the above paper, with $x_1$= 0.33 for $M <
1M_{\odot}$ and $x_2$=0.95 for $M>1M_{\odot}$.

\end{itemize}

\section{Acknowledgements}

We thank Cristina Chiappini and Andy McWilliam for many enlightening 
discussions. 
We also thank the referee Dante Minniti for his careful reading and useful suggestions.
We acknowledge financial support from PRIN2007-MIUR 
(Italian Ministry of  University and Research) Prot.2007JJC53X-001.


\begin{thebibliography}{1000}

\bibitem{Alves}
Alves-Brito, A., Mel\'endez, J., Asplund, M., Ram\'irez, I., Yong, D.
2010, A\&A, 513, 35

\bibitem{BKS}
Ballero, S.K., Kroupa, P., Matteucci, F.  2007b A\&A, 467, 117

\bibitem{b10}
Ballero, S.K., Matteucci, F., Origlia, L., Rich, R.M. 
2007a, A\&A, 467, 123

\bibitem{BensbyFeltzing2010}
Bensby, T., Feltzing, S., Johnson, J.A., et al. 
2010, A\&A, 512, 41

\bibitem{cescutti08}
Cescutti, G. 
2008, A\&A, 481, 691

\bibitem{Cescutti2009}
Cescutti, G., Matteucci, F., McWilliam, A., Chiappini, C.
2009, A\&A, 505, 605

\bibitem{b100} 
  Cescutti, G., Fran\c cois, P., Matteucci, F., Cayrel, R., Spite, M.,
  2006, \aap, 448, 557

\bibitem{Elmegreen}
Elmegreen, B.G. 2009, in Galaxy Evolution: Emerging Insights and
New Challenges, ed. S. Jogee, L. Hao, G. Blanc, \& I. Marinova,
ASP Conference Series, Vol. 419, 23

\bibitem{Elmegreen2}
Elmegreen, B.G., Bournaud, F., Elmegreen, D.M. 
2008, ApJ, 688, 67

\bibitem{b30}
Fran\c cois, P., Matteucci, F., Cayrel, R., et al.
2004, A\&A, 421, 613

\bibitem{Fulbright}
Fulbright, J.P., McWilliam, A., \& Rich, R.M. 
2007 ApJ, 661, 1152

\bibitem{greggio}
Greggio, L., Renzini, A. 
1983 A\&A, 118, 217

\bibitem{Iwamoto}
Iwamoto, K., Brachwitz, F., Nomoto, K., et al.
1999, ApJS, 125, 439

\bibitem {Johnson1}	
Johnson, J.A., Gal-Yam, A., Leonard, D.C., et al.,
2007, ApJ, 655, 33

\bibitem {Johnson2}	
Johnson, J.A., Gaudi, B.S., Sumi, T., et al.
2008, ApJ, 685, 508

\bibitem{Kodama}
Kodama, T. 1997, PhD thesis, University of Tokio

\bibitem{kroupa}
Kroupa, P. 2001, MNRAS 332, 231

\bibitem{kroupa2}
Kroupa, P., Tout, C. A., Gilmore, G.
1993, MNRAS, 262, 545

\bibitem{Lanfranchi3}
Lanfranchi, G.A., Matteucci, F., Cescutti, G. 
2006, MNRAS, 365, 477

\bibitem{Lecureur}
Lecureur, A., Hill, V., Zoccali, M., et al. 
2007, A\&A, 465, 799

\bibitem{Matteucci1}
Matteucci, F. 1992, ApJ, 397, 32

\bibitem{matteucci2}
Matteucci, F. 2001, ASSL Vol. 253, The Chemical Evolution of
the Galaxy. Kluwer, Dordrecht, 293

\bibitem{Matteucci3} 
Matteucci, F., Brocato, E. 1990, ApJ, 365, 539

\bibitem{Matteucci8}
Matteucci, F., Recchi, S. 2001, ApJ, 558, 351

\bibitem{matteucci9}
Matteucci, F., Romano, D., Molaro, P.
1999, A\&A, 341, 458

\bibitem{Maeder}
Maeder, A. 
1992, A\&A, 264, 105

\bibitem{McWilliam}	
McWilliam, A., Matteucci, F., Ballero, S.K. et al.
2008, AJ, 136, 367

\bibitem[\protect\citeauthoryear{McWilliam \& Rich}{1994}]{b325} 	
McWilliam, A., Rich, R.M.
1994, ApJS, 91, 749

\bibitem{melendez}
Mel\'endez, J., Asplund, M., Alves-Brito, A., et al. 
2008, A\&A, 484, L21

\bibitem{Meynet}
Meynet, G., \& Maeder, A. 
2002, A\&A, 390, 561

\bibitem{Minniti}
Minniti, D. 1996 ApJ, 459, 175 

\bibitem{111} 
Nomoto, K., Thielemann, F.-K., Wheeler, J.C.
1984, ApJ, 279, 23

\bibitem{Renzini2}
Renzini, A 
2006, ARA\&A, 44, 141

\bibitem{Rich05}
Rich, R.M., Origlia, L.
2005, ApJ, 634, 1293

\bibitem{Ryde09}
Ryde, N., Edvardsson, B., Gustafsson, B., et al. 2009, A\&A, 496, 701

\bibitem{Salpeter} 
Salpeter, E.E., 1955, ApJ, 121, 161

\bibitem{b37}
Scalo, J.M.
1986, FCPh, 11, 1

\bibitem{vandenHoek}
van den Hoek, L.B., \& Groenewegen, M.A.T. 
1997,  A\&AS, 123, 305 

\bibitem{b80} 
Woosley, S.E., Weaver, T.A.
1995, ApJ, 101, 181 

\bibitem{Wyse}
Wyse, R.F.G. \& Gilmore, G. 
1992, AJ, 104, 144

\bibitem{Zoccali2}
Zoccali, M., Hill, V., Lecureur, A., et al.  
2008, A\&A, 486, 177

\bibitem{Zoccali}
Zoccali, M., Lecureur, A., Barbuy, B., et al. 
2006, A\&A, 457, 1



\end{thebibliography}
\end{document}